\documentclass[twoside,amsmath,amssymb,notitlepage,superscriptaddress,showpacs]{revtex4-1}
\usepackage[latin9]{inputenc}
\usepackage{amsmath}
\usepackage{graphicx}
\usepackage{wasysym}

\makeatletter

\providecommand{\tabularnewline}{\\}

\usepackage{dcolumn}% Align table columns on decimal point
\usepackage{bm}% bold math
\makeatother

\begin{document}

\title{Dynamic structure factor of a strongly correlated Fermi superfluid
within a density functional theory approach}

\author{Peng Zou}
\email{phy.zoupeng@gmail.com}

\affiliation{Centre for Quantum and Optical Science, Swinburne University of Technology,
Melbourne 3122, Australia}

\author{Franco Dalfovo}

\affiliation{INO-CNR BEC Center and Dipartimento di Fisica, Universit\`a di Trento,
38123 Povo, Italy}

\author{Rishi Sharma}

\affiliation{Department of Theoretical Physics, Tata Institute of Fundamental
Research, Homi Bhabha Road, Mumbai 400005, India}

\author{Xia-Ji Liu}

\affiliation{Centre for Quantum and Optical Science, Swinburne University of Technology,
Melbourne 3122, Australia}

\author{Hui Hu}

\affiliation{Centre for Quantum and Optical Science, Swinburne University of Technology,
Melbourne 3122, Australia}

\date{\today}
\begin{abstract}
We theoretically investigate the dynamic structure factor of a strongly
interacting Fermi gas at the crossover from Bardeen-Cooper-Schrieffer
superfluids to Bose-Einstein condensates, by developing an improved
random phase approximation within the framework of a density functional
theory - the so-called superfluid local density approximation. Compared
with the previous random-phase-approximation studies based on the
standard Bogoliubov-de Gennes equations, the use of the density functional
theory greatly improves the accuracy of the equation of state at the
crossover, and leads to a better description of both collective Bogoliubov-Anderson-Goldstone
phonon mode and single-particle fermionic excitations at small transferred
momentum. Near unitarity, where the $s$-wave scattering length diverges,
we show that the single-particle excitations start to significantly
contribute to the spectrum of dynamic structure factor once the frequency
is above a threshold of the energy gap at $2\Delta$. The sharp rise
in the spectrum at this threshold can be utilized to measure the pairing
gap $\Delta$. Together with the sound velocity determined from the
phonon branch, the dynamic structure factor provides us some key information
of the crossover Fermi superfluid. Our predictions could be examined
in experiments with $^{6}$Li or $^{40}$K atoms using Bragg spectroscopy.
\end{abstract}

\maketitle

\section{Introduction}

The realization of ultracold Fermi gases of $^{6}$Li and $^{40}$K
atoms near Feshbach resonances provides a new paradigm for studying
strongly correlated many-body systems \cite{Bloch2008}. At low temperature,
these systems display an intriguing crossover from Bardeen-Cooper-Schrieffer
(BCS) superfluids to Bose-Einstein condensates (BEC)\cite{Leggett1980,Giorgini2008}.
At a special point in between the two regimes, where the s-wave scattering
length diverges, the gas exhibits universal properties, which might
also exist in other strongly interacting Fermi superfluids \cite{Ho2004,Hu2007},
such as high-temperature superconductors or nuclear matter in neutron
stars. This is called unitary Fermi gas and corresponds a novel type
of superfluid with neither dominant bosonic nor fermionic character.
This new superfluid has already been intensively investigated \cite{Giorgini2008},
leading to several milestone observations.

Theoretical challenges in describing the BCS-BEC crossover arise from
its strongly correlated nature: there is no small interaction parameter
to control the accuracy of theories \cite{Hu2010}. To date, significant
progress has been made in developing better quantum Monte Carlo (QMC)
simulations \cite{Astrakharchik2004,Bulgac2006,Burovski2008,Carlson2005,Carlson2008,Carlson2011,Forbes2011,Gandolfi2014,Carlson2014}
and strong-coupling theories \cite{Ohashi2003,Liu2005,Chen2005,Hu2006,Haussmann2007,Diener2008,Mulkerin2016},
leading to the quantitative establishment of a number of properties
in conjunction with the rapid experimental advances. These include
the equation of state \cite{Hu2010,Luo2007,Nascimbene2010,Horikoshi2010,Navon2010,Ku2012},
frequency of collective oscillations \cite{Hu2004,Altmeyer2007},
pairing gap \cite{Carlson2005,Carlson2008,Schunck2007,Schirotzek2008},
and superfluid transition temperature \cite{Burovski2008,Ku2012}.
However, some fundamental dynamical properties, such as the single-particle
spectral function measured by radio-frequency (rf) spectroscopy \cite{Gaebler2010,Massignan2008,Chen2009,Hu2010PRL}
and the dynamic structure factor probed by Bragg spectroscopy \cite{Combescot2006,Veeravalli2008,Hu2012,Lingham2014,Vale2016},
are not well understood yet.

As an important fingerprint of quantum gases in some certain states,
dynamic structure factor contains rich information of properties of
a many-body system \cite{Pitaevskii2003}. By tuning the transferred
momentum or energy from a low value to a high one, we can observe
the low-lying collective phonon excitations, Cooper-pair (i.e., molecular)
excitations and single-particle atomic excitations, respectively.
In particular, at finite temperature the dynamic structure factor
can help to judge whether the system is in the superfluid or normal
state from the emergence of the phonon excitations. Also, it is reasonable
to anticipate that the dynamic structure factor may play a role to
solve the debate on the existence of pseudogap pairing or pre-pairing
states \cite{Massignan2008,Chen2009}. Experimentally, the dynamic
structure factor can be measured via two-photon Bragg scattering technique
\cite{Veeravalli2008}, at both low and finite temperatures \cite{Lingham2014}.
Theoretically, since no exact solution exists for strongly interacting
Fermi gases and the numerically exact QMC approach is less efficient
for simulating dynamical quantities, one has to resort to some approximated
approaches, which are useful in certain limiting cases \cite{Hu2012}
(see Table I). For example, at high temperature, as the fugacity is
a small parameter, a quantum cluster expansion has been proven to
be an efficient method \cite{Liu2009PRL,Liu2013}, and has been used
to calculate the dynamic structure factor \cite{Hu2010PRA,Shen2013}.
In the limits of both large momentum and high frequency, asymptotically
exact Tan relations have been derived to describe the high-frequency
tails \cite{Son2010,Hu2012PRA}. On the other hand, in the limit of
long wavelength or small momentum, the phenomenological two-fluid
hydrodynamic theory may provide a useful description \cite{Hu2010NJP}.

\begin{table}
\begin{centering}
\begin{tabular}{|c|c|c|}
\hline 
Theories &  $q$ (applicable) & $T$ (applicable)\tabularnewline
\hline 
\hline 
Virial expansion & arbitrary & $T>T_{F}$\tabularnewline
\hline 
Tan relation & $q\gg k_{F}$, $\omega\gg\varepsilon_{F}$ & $T\gg T_{F}$\tabularnewline
\hline 
Two-fluid hydrodynamics & $q\ll k_{F}$,  & $T<T_{c}$\tabularnewline
\hline 
Diagrammatic approach & arbitrary & arbitrary\tabularnewline
\hline 
BdG-RPA & $q\gg k_{F}$ & $T\ll T_{c}$\tabularnewline
\hline 
SLDA-RPA (this work) & $q\lesssim k_{F}$ & $T\ll T_{c}$\tabularnewline
\hline 
\end{tabular}
\par\end{centering}
\caption{A list of the existing theories for the dynamic structure of strongly
interacting fermions, including the virial expansion \cite{Hu2010PRA,Shen2013},
Tan relation \cite{Son2010,Hu2012PRA}, two-fluid hydrodynamics \cite{Hu2010NJP},
diagrammatic strong-coupling approach \cite{Palestini2012,He2016},
and BdG-RPA \cite{Combescot2006,Combescot2006EPL,Zou2010,Guo2013}.
The applicable conditions for the transferred momentum $q$ and temperature
$T$, under which each theory is quantitatively useful, are indicated.
Here $k_{F}=(3\pi^{2}n)^{1/3}$, $\varepsilon_{F}=k_{F}^{2}/(2m)=(3\pi^{2}n)^{2/3}/(2m),$
and $T_{F}$ are the Fermi momentum, energy, and temperature, respectively.
$T_{c}$ is the superfluid transition temperature.}

\label{table1} 
\end{table}

A general theoretical framework of the dynamic structure factor, valid
at \emph{arbitrary} temperature and momentum, can be developed by
using the diagrammatic technique \cite{Guo2010,Palestini2012} or
functional path integral approach \cite{He2016}, in parallel with
the existing strong-coupling theories of interacting Fermi gases \cite{Hu2010}.
The expressions for the density and spin responses of strongly interacting
Fermi gases have been obtained \cite{He2016}. However, their numerical
calculations turn out to be extremely difficult, except in the limit
of zero transferred momentum \cite{Palestini2012}. A more commonly
used approach is the random-phase approximation (RPA) on top of the
mean-field Bogoliubov-de Gennes (BdG) theory \cite{Combescot2006,Combescot2006EPL,Zou2010,Guo2013}.
By comparing the BdG-RPA predictions with the experimental data for
the dynamic structure factor of strongly interacting fermions and
with the QMC results for the static structure factor \cite{Zou2010},
it has been surprisingly shown by two of the present authors that
the BdG-RPA theory works \emph{quantitatively} well at sufficiently
large transferred momentum (i.e., $q\sim5k_{F}$, where $k_{F}$ is
the Fermi momentum). At small transferred momentum, i.e., $q\lesssim k_{F}$,
apparently, the BdG-RPA only provides a qualitative description of
the dynamic structure factor at the BCS-BEC crossover, since both
the sound velocity (associated with the phonon excitations) and pairing
gap (associated with the single-particle fermionic excitations) are
strongly over-estimated within the BdG framework \cite{Giorgini2008}.

In this work, we aim to develop a \emph{quantitative} theory for the
dynamic structure factor of strongly interacting fermions at low transferred
momentum and at low temperature, which is amenable for numerical calculations.
For this purpose, we adopt a superfluid local density approximation
(SLDA) approach \cite{Bulgac2002,Yu2003,Bulgac2007}, within the framework
of density functional theory \cite{Hohenberg1964,Kohn1965,Kohn1999},
as recently suggested by Bulgac and his co-workers. The SLDA theory
assumes an energy density functional (i.e., a function of the density
function) to describe a unitary Fermi superfluid and uses the QMC
results for the chemical potential and order parameter as two important
inputs. It can be well regarded as a better quasi-particle description
than the mean-field BdG theory. It has been shown that at low-energy
the SLDA theory provides useful results for the equation of state
\cite{Bulgac2007} and real-time dynamics \cite{Bulgac2011,Bulgac2013}
of a strongly interacting Fermi superfluid.

Here we apply the random phase approximation on top of the SLDA theory.
The use of SLDA in place of the standard BdG equations improves the
predictions for the dynamic structure factor in the BCS-BEC crossover
near unitarity. The static structure factor at small momentum transfer
is in excellent agreement with the results of the latest QMC \cite{Gandolfi2014,Carlson2014,Combescot2006EPL}.
A more stringent test can be obtained in the near future by comparing
our predictions with the experimental data \cite{Vale2016}, without
any adjustable parameters.

Our paper is organized as follows. In the next section (Sec. \ref{slda}),
we introduce the SLDA theory. In Sec. \ref{rpa}, we review the main
idea of RPA. The expression for the dynamic structure factor is derived
in Sec. \ref{dsf}. In Sec. \ref{unitaryfermigas} and Sec. \ref{cross},
we present our main results of dynamic structure factor in the unitary
limit and the crossover regime, respectively. Finally, Sec. \ref{conclusions}
is devoted to conclusions and outlooks. For convenience, we set $\hbar=k_{B}=1$
in the following discussions.

\section{Superfluid local density approximation}

\label{slda}

The density functional theory (DFT) developed by Hohenberg and Kohn
\cite{Hohenberg1964}, together with the local density approximation
(LDA) by Kohn and Sham \cite{Kohn1965}, is a powerful tool to understand
the properties of many-electron systems. The DFT was initially used
for electrons in the normal, non-superconducting state. It is based
on the assumptions that there is a unique mapping between the external
potential and the total wave function of the system (or the normal
density), and that the exact energy of the system can be written as
a density functional. A limitation of the DFT is that the exact form
of the density functional is often not known. Therefore, approximated
phenomenological functionals are introduced, which should be optimized
for a specific system. Typically, those functionals rely on the Kohn-Sham
orbitals \cite{Kohn1999} and thus can not effectively deal with superfluidity.
The generalization of the DFT to superfluid cold-atom systems - referred
to as SLDA as we mentioned earlier - was recently introduced by Bulgac
and Yu \cite{Bulgac2002,Yu2003,Bulgac2007}. This SLDA originates
from a similar DFT previously used in the context of nuclear physics
\cite{Yu2003,Bulgac2013}.

A nice feature of ultracold fermions is that, in the unitary limit
the form of the energy density functional is restricted by dimensional
arguments. Another advantage is the availability of \emph{ab-initio}
QMC results and accurate experimental data for both homogeneous and
inhomogeneous systems, which can be used to fix the parameters of
the density functional, as we shall see below.

For a superfluid atomic Fermi gas, two atoms with mass $m$ in different
spin state can form a Cooper pair. As a result, the system possesses
an anomalous Cooper-pair density $\nu(\mathbf{r},t)$, in addition
to the number density $n(\mathbf{r},t)$. The energy density functional
$\mathcal{E}[\tau(\mathbf{r},t),n(\mathbf{r},t),\nu(\mathbf{r},t)]$
of the system must include the kinetic density $\tau(\mathbf{r},t)$,
number density $n(\mathbf{r},t)$, and also the anomalous density
$\nu(\mathbf{r},t)$ \cite{Yu2003,Bulgac2007}: 
\begin{equation}
\mathcal{E}\left[\tau,n,\nu\right]=\alpha\frac{1}{2m}\tau+\beta\frac{3(3\pi^{2})^{2/3}}{10m}n^{5/3}+\gamma\frac{1}{mn^{1/3}}|\nu|^{2},\label{edslda}
\end{equation}
where the kinetic density $\tau$, number density $n$ and anomalous
density $\nu$ are given by, 
\begin{equation}
\tau=2\sum_{\mathbf{k}}|\nabla v_{\mathbf{k}}|^{2},\:n=2\sum_{\mathbf{k}}|v_{\mathbf{k}}|^{2},\:\nu=\sum_{\mathbf{k}}u_{\mathbf{k}}v_{\mathbf{k}}^{*},\label{den3}
\end{equation}
and $u_{\mathbf{k}}(\mathbf{r},t)$ and $v_{\mathbf{k}}(\mathbf{r},t)$
are the Bogoliubov quasiparticle wavefunctions with $\mathbf{k}$
labeling the quasiparticle states. Three dimensionless constants,
the effective mass parameter $\alpha$, Hartree parameter $\beta$
and pairing parameter $\gamma$, are introduced. These parameters
are determined by requiring that the SLDA reproduces exactly the \emph{zero
temperature} chemical potential, pairing gap and energy per particle
that are obtained by either QMC simulations or accurate experimental
measurements for a uniform system \cite{Bulgac2007,Bulgac2013}. 

In the unitary limit at zero temperature, the simple form of the energy
density functional Eq. (\ref{edslda}) is inspired by the dimensional
analysis: the first and third terms are the unique combination required
by the renormalizablity of the theory \cite{Bulgac2002}; while the
second term is the only possible form allowed by the scale invariance
at unitarity \cite{Ho2004}. The above energy density functional has
been successfully used by Bulgac and his co-workers to understand
the thermodynamics \cite{Bulgac2007} and dynamics \cite{Bulgac2011,Bulgac2013}
of a unitary Fermi gas at zero temperature. It is reasonable to assume
that the energy density functional Eq. (\ref{edslda}) can be applied
also away from unitarity, but close to it, and at non-zero temperature,
but significantly below $T_{c}$.

As both the kinetic and anomalous densities diverge due to the use
of a pairwise contact interaction, a regularization procedure is needed
for the pairing gap and for the energy density \cite{Bulgac2002}.
After regularization, the energy density functional with regularized
kinetic density $\tau_{c}(\mathbf{r},t)$ and anomalous density $\nu_{c}(\mathbf{r},t)$
takes the following form \cite{Bulgac2007},

\begin{equation}
\mathcal{E}=\alpha\frac{1}{2m}\tau_{c}+\beta\frac{3(3\pi^{2})^{2/3}}{10m}n^{5/3}+g_{{\rm eff}}|\nu_{c}|^{2},\label{edsldarg}
\end{equation}
where the effective coupling constant $g_{{\rm eff}}$ is given by
\begin{equation}
\frac{1}{g_{{\rm eff}}}=\frac{mn^{1/3}}{\gamma}-\sum_{\mathbf{\left|k\right|<\Lambda}}\frac{m}{\alpha\mathbf{k}^{2}}.\label{sldageff}
\end{equation}
We note that, $g_{\textrm{eff}}$ scales to zero once the cut-off
momentum $\Lambda$ runs to infinity. The order parameter $\Delta(\mathbf{r},t)$
is related to the anomalous density $\nu$ by 
\begin{equation}
\Delta\left(\mathbf{r},t\right)=-g_{{\rm eff}}\nu_{c}\left(\mathbf{r},t\right).\label{delta}
\end{equation}
The stationary SLDA equations for the quasiparticle wave functions
are obtained by the standard functional minimization with respect
to the variations $u_{\mathbf{k}}$ and $v_{\mathbf{k}}$. One obtains
\begin{equation}
\left[\begin{array}{cc}
\mathcal{H}_{s}-\mu & \Delta\\
\Delta^{*} & -\mathcal{H}_{s}+\mu
\end{array}\right]\left[\begin{array}{c}
u_{\mathbf{k}}\\
v_{\mathbf{k}}
\end{array}\right]=E_{\mathbf{k}}\left[\begin{array}{c}
u_{\mathbf{k}}\\
v_{\mathbf{k}}
\end{array}\right],\label{bdgslda}
\end{equation}
with a single quasiparticle Hamiltonian 
\begin{equation}
\begin{split}\mathcal{H}_{s} & =-\alpha\frac{\nabla^{2}}{2m}+\beta\frac{\left(3\pi^{2}n\right){}^{2/3}}{2m}-\frac{|\Delta|^{2}}{3\gamma mn^{2/3}},\end{split}
\end{equation}
and the chemical potential $\mu$.

By requiring that a homogeneous Fermi gas of the number density $n=N/V=k_{F}^{3}/(3\pi^{2})$
has an energy per particle $E/N=(3/5)\xi_{E}\varepsilon_{F}$, a chemical
potential $\mu=\xi_{\mu}\varepsilon_{F}$, and a pairing order parameter
$\Delta=\eta\varepsilon_{F}$ at zero temperature, one can determine
the value of dimensionless parameters $\alpha$, $\beta$ and $\gamma$
in Eq. (\ref{edsldarg}) through the following equations, which are
independent on the cut-off momentum (i.e., $\Lambda\rightarrow\infty$):
\begin{equation}
n=\sum_{\mathbf{k}}\left(1-\frac{\xi_{\mathbf{k}}}{E_{\mathbf{k}}}\right),\label{3ep1}
\end{equation}

\begin{equation}
\frac{mn^{1/3}}{\gamma}=\sum_{\mathbf{k}}\left(\frac{m}{\alpha\mathbf{k}^{2}}-\frac{1}{2E_{\mathbf{k}}}\right),\label{3ep2}
\end{equation}
and
\begin{equation}
\frac{3}{5}E_{F}n(\xi_{E}-\beta)=\sum_{\mathbf{k}}\left[\alpha\frac{\mathbf{k}^{2}}{2m}\left(1-\frac{\xi_{\mathbf{k}}}{E_{\mathbf{k}}}\right)-\frac{\Delta}{2E_{\mathbf{k}}}\right],\label{3ep3}
\end{equation}
where $\xi_{\mathbf{k}}=\alpha\mathbf{k}^{2}/(2m)+[\beta-(3\pi^{2})^{2/3}\eta^{2}/(6\gamma)-\xi_{\mu}]\varepsilon_{F}$
and $E_{\mathbf{k}}=\sqrt{\xi_{\mathbf{k}}^{2}+\Delta^{2}}$. In these
three constraint equations, $\xi_{\mu}$, $\eta$ and $\xi_{E}$ are
the three inputs, whose value can be reliably determined by using
QMC simulations \cite{Carlson2008,Carlson2011,Forbes2011} or from
the experimental measurements \cite{Navon2010,Ku2012,Schunck2007}.
The parameter $\alpha$ can be determined using the single particle
dispersion, near the unitary limit, typically the parameter $\alpha$
is very close to $1$ \cite{Bulgac2007,Zwerger2012}, indicating that
the effective mass only differs slightly from the bare atomic mass
$m$. For simplicity, throughout the work we take $\alpha=1$ and
use the density equation Eq. (\ref{3ep1}) and the gap equation Eq.
(\ref{3ep2}) to determine the parameters $\beta$ and $\gamma$.
As we shall see, this simple choice also ensures that the $f$-sum
rule of the dynamic structure factor is strictly satisfied.

For a unitary Fermi superfluid, where $\xi_{\mu}=\xi_{E}=\xi$ due
to the scale invariance \cite{Ho2004}, the latest auxiliary field
QMC provides $\xi\simeq0.372$ \cite{Carlson2011}, which is quite
close to the experimental value $\xi=0.376(5)$ \cite{Ku2012}. As
to the parameter $\eta$, its accurate value is to be determined yet.
An earlier rf-spectroscopy experiment reports $\eta\simeq0.44$ \cite{Schirotzek2008}
and the latest QMC result is $\eta=0.504$ \cite{Carlson2005,Carlson2008}.
In this work, for a unitary Fermi gas we choose the experimental result
$\mu=0.376\varepsilon_{F}$ for the chemical potential and the QMC
prediction $\Delta=0.5\varepsilon_{F}$ for the pairing gap. This
leads to $\beta=-0.430$ and $1/\gamma=-0.0767$. It is worth noting
that, when $\alpha=1$, our SLDA result reduces that of the standard
BdG theory, if we set $\beta$ and $1/\gamma$ to zero.

Away from the unitary limit, the knowledge on the pairing gap is not
complete. We use the predictions of a Gaussian pair fluctuation theory
\cite{Hu2006,Diener2008} as the inputs, since these theoretical results
have already been shown to provide a satisfactory explanation for
the experimentally measured chemical potential \cite{Navon2010}. 

\section{Random phase approximation}

\label{rpa}

If a superfluid Fermi gas is perturbed by a small external potential,
usually the number density and anomalous density will fluctuate. Due
to the interatomic interactions, the fluctuating densities will feedback
and induce an additional perturbation potential. One way to include
these fluctuation effects is to use the linear response theory within
the RPA \cite{Combescot2006,Combescot2006EPL,Minguzzi2001,Bruun2001,Liu2004,Stringari2009}.
The essential idea of RPA is that the induced fluctuation potential
is assumed to be a \emph{self-generated} mean-field potential experienced
by quasiparticles, due to the local changes in the number densities
$n_{\uparrow}(\mathbf{r},t)$ and $n_{\downarrow}(\mathbf{r},t)$,
and Cooper-pairs density $\nu(\mathbf{r},t)$ or its complex conjugate
$\nu^{*}(\mathbf{r},t)$. In the following, for convenience, we denote
these four densities $n_{\uparrow}$, $n_{\downarrow}$, $\nu$ and
$\nu^{*}$ as $n_{1}$, $n_{2}$, $n_{3}$ and $n_{4}$, respectively.

In the SLDA energy density functional, it is easy to see that the
interaction contribution to the functional is given by,
\begin{equation}
\mathcal{E}_{{\rm int}}=\beta\frac{3(3\pi^{2})^{2/3}}{10m}n^{5/3}\left(\mathbf{r},t\right)+\frac{|\Delta\left(\mathbf{r},t\right)|^{2}}{g_{{\rm eff}}}.\label{eint}
\end{equation}
The resulting fluctuating potential is simply $\sum_{j}E_{ij}^{I}\delta n_{j}$,
where \cite{Stringari2009} 
\begin{equation}
E_{ij}^{I}=\left(\frac{\partial^{2}\mathcal{E}_{{\rm int}}}{\partial n_{i}\partial n_{j}}\right)_{0}
\end{equation}
and $\delta n_{i=1,2,3,4}$ are the density fluctuations around equilibrium,
which are to be determined. The suffix $0$ indicates that the derivatives
are calculated at equilibrium. Therefore, together with the external
potential $V_{{\rm ext}}^{i}$, the total effective perturbative potential
takes the form, 
\begin{equation}
V_{{\rm eff}}^{i}\equiv V_{{\rm ext}}^{i}+\underset{j}{\sum}E_{ij}^{I}\delta n_{j}.\label{veff}
\end{equation}
Using this effective perturbation, the density fluctuations $\delta n_{i}$
can be written down straightforwardly, according to the standard linear
response theory, 
\begin{equation}
\delta n_{i}=\underset{j}{\sum}\chi_{ij}^{0}V_{{\rm eff}}^{i},\label{neff}
\end{equation}
where $\chi^{0}$ is the \emph{bare} response function of the quasiparticle
reference system described by the SLDA equation (\ref{bdgslda}),
which is easy to calculate (see Appendix A). By combining Eqs. (\ref{veff})
and (\ref{neff}), we arrive at, 
\begin{equation}
\delta n_{i}=\underset{j}{\sum}\chi_{ij}V_{{\rm ext}}^{i},\label{next}
\end{equation}
where $\chi$ is the RPA response function, 
\begin{equation}
\chi=\chi^{0}\left[1-\chi^{0}E^{I}\right]^{-1}.\label{kappa}
\end{equation}
Once the bare response function $\chi^{0}$ and the second order derivative
$E_{ij}^{I}$ are known, we obtain directly $\chi$. The density response
function $\chi_{D}$ is a summation of $\chi_{ij}$ in the density
channel: $\chi_{D}=\chi_{11}+\chi_{12}+\chi_{21}+\chi_{22}=2(\chi_{11}+\chi_{12})$.
The dynamic structure factor is connected to the imaginary part of
the density response function,

\begin{equation}
S(\mathbf{q},\omega)=-\frac{1}{\pi}\frac{{\rm Im}\chi_{D}\left(\mathbf{q},i\nu_{n}\rightarrow\omega+i0^{+}\right)}{1-e^{-\omega/T}},\label{sqw}
\end{equation}
with $\mathbf{q}$ and $\omega$ being the transferred momentum and
energy, respectively.

The RPA on top of the mean-field BdG theory has previously been used
to study the dynamic structure factor \cite{Minguzzi2001} and collective
oscillations \cite{Bruun2001} of weakly interacting Fermi superfluids.
A dynamical mean-field approach, identical to the RPA but based on
kinetic equations, was also developed to investigate dynamic and static
structure factors and collective modes of strongly interacting Fermi
superfluids \cite{Combescot2006,Combescot2006EPL}. Some properties
of the density response of unitary Fermi gas for the SLDA has also
been studied in \cite{Forbes2014}. In the following, we examine the
improved RPA based on the SLDA theory.

\section{Dynamic structure factor in SLDA theory}

\label{dsf}

The calculation of the second-order derivative matrix $E^{I}$ is
straightforward. It reads, 
\begin{equation}
E^{I}=\left[\begin{array}{cccc}
\mathcal{I}_{n}\varepsilon_{F}/n & \mathcal{I}_{n}\varepsilon_{F}/n & \mathcal{I}_{\nu}g_{e\textrm{ff}} & \mathcal{I}_{\nu}g_{e\textrm{ff}}\\
\mathcal{I}_{n}\varepsilon_{F}/n & \mathcal{I}_{n}\varepsilon_{F}/n & \mathcal{I}_{\nu}g_{e\textrm{ff}} & \mathcal{I}_{\nu}g_{e\textrm{ff}}\\
\mathcal{I}_{\nu}g_{e\textrm{ff}} & \mathcal{I}_{\nu}g_{e\textrm{ff}} & 0 & g_{{\rm eff}}\\
\mathcal{I}_{\nu}g_{e\textrm{ff}} & \mathcal{I}_{\nu}g_{e\textrm{ff}} & g_{{\rm eff}} & 0
\end{array}\right],
\end{equation}
where $\mathcal{I}_{n}$ and $\mathcal{I}_{\nu}$ are two dimensionless
variables,
\begin{eqnarray*}
\mathcal{I}_{n} & = & \frac{2\beta}{3}+\frac{\left(3\pi^{2}\right)^{2/3}}{9\gamma}\frac{\Delta{}^{2}}{\varepsilon_{F}^{2}},\\
\mathcal{I}_{\nu} & = & \frac{\left(3\pi^{2}\right)^{2/3}}{6\gamma}\frac{\Delta}{\varepsilon_{F}}.
\end{eqnarray*}
We note the existence of the crossing term $\mathcal{I}_{\ensuremath{\nu}}$,
due to the (implicit) coupling between the number density and the
anomalous Cooper-pair density in the interaction energy density functional
Eq. (\ref{eint}). In the unitary limit, in comparison to the BdG-RPA
theory, we note also that the matrix element in the number density
channel, $\mathcal{I}_{n}\varepsilon_{F}/n$, changes from a vanishingly
small number (i.e., at the order of $g_{{\rm eff}}$) to a finite
value. The response function of the quasiparticle reference system
$\chi^{0}$ can be constructed by solving the stationary SLDA equation
(\ref{bdgslda}). It is a $4$ by $4$ matrix. However, as we shown
in Appendix A, only six of all 16 matrix elements are independent:

\begin{equation}
\chi^{0}=\left[\begin{array}{cccc}
\chi_{11}^{0} & \chi_{12}^{0} & \chi_{13}^{0} & \chi_{14}^{0}\\
\chi_{12}^{0} & \chi_{11}^{0} & \chi_{13}^{0} & \chi_{14}^{0}\\
\chi_{14}^{0} & \chi_{14}^{0} & -\chi_{12}^{0} & \chi_{34}^{0}\\
\chi_{13}^{0} & \chi_{13}^{0} & \chi_{43}^{0} & -\chi_{12}^{0}
\end{array}\right],\label{kappa0}
\end{equation}
The detailed expressions of the elements $\chi_{11}^{0}$, $\chi_{12}^{0}$,
$\chi_{13}^{0}$, $\chi_{14}^{0}$, $\chi_{34}^{0}$ and $\chi_{43}^{0}$
are we show in Appendix A. By solving the RPA equation (\ref{kappa}),
we obtain all the matrix elements $\chi_{ij}$ of the RPA response
function $\chi$. The resulting density response function is given
by, 
\begin{equation}
\chi_{D}=2\left|\begin{array}{ccc}
\chi_{12}^{0}+\chi_{11}^{0} & -\chi_{14}^{0}g_{{\rm eff}} & -\chi_{13}^{0}g_{{\rm eff}}\\
2\chi_{14}^{0} & 1-\chi_{34}^{0}g_{{\rm eff}} & \chi_{12}^{0}g_{{\rm eff}}\\
2\chi_{13}^{0} & \chi_{12}^{0}g_{{\rm eff}} & 1-\chi_{43}^{0}g_{{\rm eff}}
\end{array}\right|/|1-\chi^{0}E^{I}|.
\end{equation}

It is well known that the anomalous density correlated functions,
$\chi_{34}^{0}$ and $\chi_{43}^{0}$, are divergent, because of the
use of the contact interatomic interaction \cite{Bruun2001}. Thus,
we introduce the regularized functions $\widetilde{\chi}_{34}^{0}=\chi_{34}^{0}-1/g_{{\rm eff}}$
and $\widetilde{\chi}_{43}^{0}=\chi_{43}^{0}-1/g_{{\rm eff}}$, with
which the density response function now takes the form, 
\begin{equation}
\chi_{D}=2\frac{\left(\mathcal{B}_{n1}-2\mathcal{B}_{n2}\right)}{\left|1-\chi^{0}E^{I}\right|/g_{{\rm eff}}^{2}}.\label{densitykappa}
\end{equation}
Here, 
\begin{eqnarray*}
\mathcal{B}_{n1} & = & \left(\chi_{11}^{0}+\chi_{12}^{0}\right)\left[\widetilde{\chi}_{34}^{0}\widetilde{\chi}_{43}^{0}-\left(\chi_{12}^{0}\right)^{2}\right],\\
\mathcal{B}_{n2} & = & 2\chi_{12}^{0}\chi_{13}^{0}\chi_{14}^{0}+\left(\chi_{13}^{0}\right)^{2}\widetilde{\chi}_{34}^{0}+\left(\chi_{14}^{0}\right)^{2}\widetilde{\chi}_{43}^{0}.
\end{eqnarray*}
To obtain the expression of $\left|1-\chi^{0}E^{I}\right|/g_{{\rm eff}}^{2}$,
it should be noted that $g_{{\rm eff}}$ is a vanishingly small quantity.
Therefore, it is useful to arrange different terms in terms of the
powers of $g_{e\textrm{ff}}$. For instance, for the matrix elements
of $E_{n}$, $\mathcal{I}_{n}\varepsilon_{F}/n$ has the order of
$[g_{{\rm eff}}]^{0}$, while $\mathcal{I}_{\nu}g_{\textrm{eff}}$
has the order of $[g_{{\rm eff}}]^{1}$. For the determinant $\left|1-\chi^{0}E^{I}\right|$,
there are no terms at the order of $\mathcal{O}(g_{{\rm eff}})$ or
$\mathcal{O}(1)$, as anticipated. The order of most terms is $\mathcal{O}([g_{{\rm eff}}]^{2})$.
By collecting those terms, we find that,

\begin{equation}
\frac{\left|1-\chi^{0}E^{I}\right|}{g_{{\rm eff}}^{2}}=4\mathcal{I}_{\nu}\mathcal{B}_{\nu1}+2\mathcal{I}_{\nu}^{2}\mathcal{B}_{\nu2}-2\mathcal{I}_{n}\left(\mathcal{B}_{n1}-2\mathcal{B}_{n2}\right)+\widetilde{\chi}_{34}^{0}\widetilde{\chi}_{43}^{0}-\left(\chi_{12}^{0}\right)^{2},\label{det}
\end{equation}

where 
\begin{eqnarray*}
\mathcal{B}_{\nu1} & = & \chi_{12}^{0}\chi_{13}^{0}+\chi_{12}^{0}\chi_{14}^{0}+\chi_{13}^{0}\widetilde{\chi}_{34}^{0}+\chi_{14}^{0}\widetilde{\chi}_{43}^{0},\\
\mathcal{B}_{\nu2} & = & (\chi_{11}^{0}+\chi_{12}^{0})\left(2\chi_{12}^{0}+\widetilde{\chi}_{34}^{0}+\widetilde{\chi}_{43}^{0}\right)-2\left(\chi_{13}^{0}-\chi_{14}^{0}\right)^{2},
\end{eqnarray*}
In the unitary limit, if we set both $\mathcal{I}_{n}$ and $\mathcal{I}_{\ensuremath{\nu}}$
to zero, $\left|1-\chi^{0}E^{I}\right|/g_{{\rm eff}}^{2}$ is just
$\widetilde{\chi}_{34}^{0}\widetilde{\chi}_{43}^{0}-\left(\chi_{12}^{0}\right)^{2}$,
and then we recover the BdG-RPA expression for the density response
function \cite{Combescot2006,Zou2010,Guo2013}. 

We use Eqs. (\ref{densitykappa}) and (\ref{det}) to obtain the density
response function $\chi_{D}$ and then calculate the dynamic structure
factor $S(\mathbf{q},\omega)$ via the fluctuation-dissipation theorem
Eq. (\ref{sqw}). To take the analytic continuation numerically, i.e.,
$i\nu_{n}\rightarrow\omega+i\delta$, where $\delta=0^{+}$, we use
a small broadening parameter $\delta=10^{-3}\varepsilon_{F}$, unless
specified elsewhere.

\section{Dynamic structure factor of a unitary Fermi superfluid}

\label{unitaryfermigas}
\begin{figure}
\begin{centering}
\includegraphics[width=0.5\textwidth]{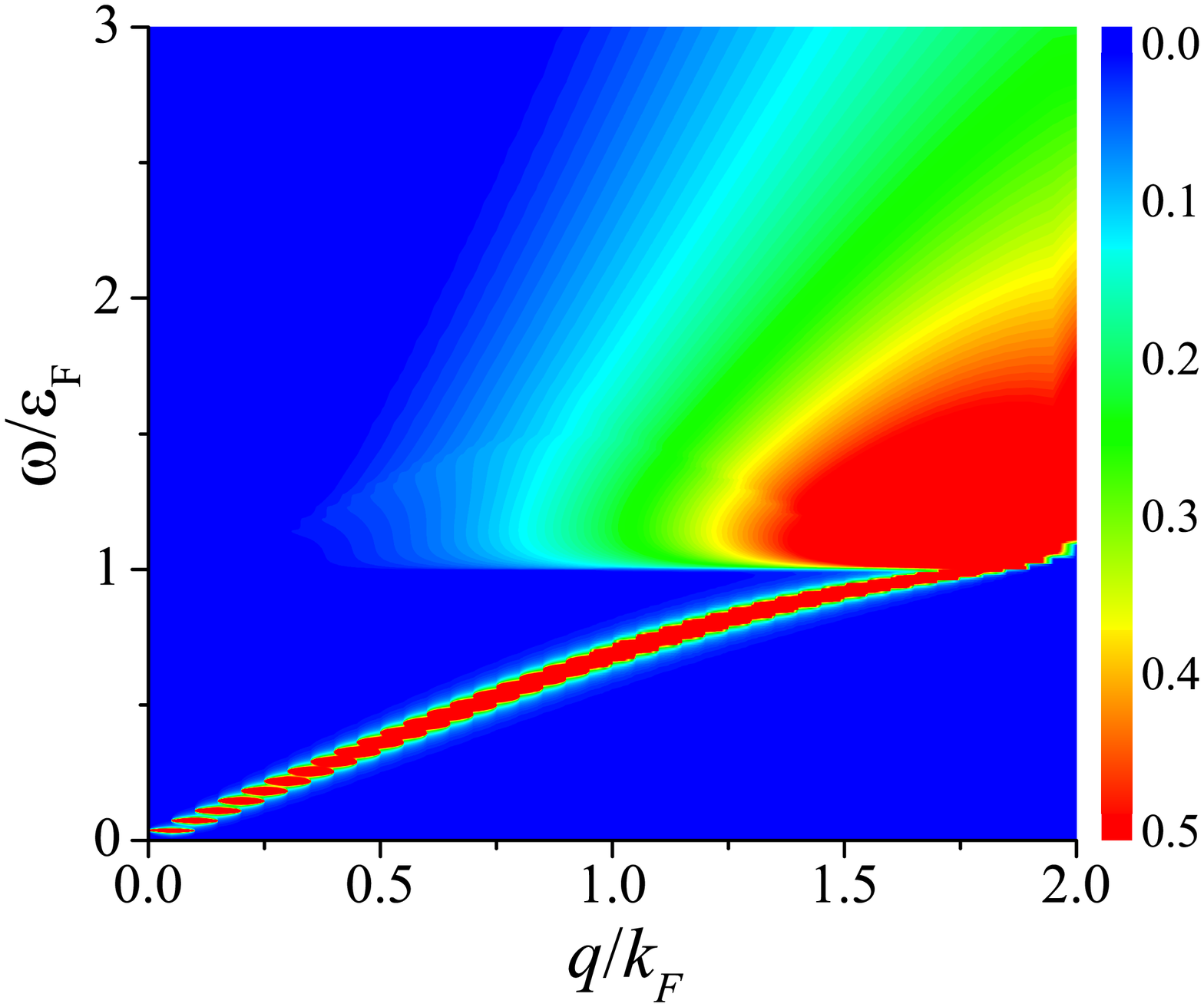} \caption{(color online). The contour plot of the dynamic structure factor of
a unitary Fermi gas at zero temperature, obtained by using SLDA-RPA.
The slope of the low-energy branch is given by the sound speed $c_{s}\simeq0.354v_{F}$,
while the horizontal threshold at $\omega\simeq\varepsilon_{F}$ is
equal to the minimum energy $2\Delta$ to break a Cooper-pair. The
color bar indicates the value of the dynamic structure factor, which
is measured in units of $N/\varepsilon_{F}$ and changes from $0$
(blue) to $0.5$ (red).}
\par\end{centering}
\label{fig1} 
\end{figure}

In this section, we present the results for the dynamic structure
factor of a unitary Fermi gas at zero temperature within SLDA-RPA,
and justify our theory at low transferred momentum $q\leq k_{F}$
by comparing the resulting static structure factor Eq. (\ref{eq:ssf})
with the latest QMC data \cite{Carlson2014}.

Fig. 1 reports a contour plot of $S(\mathbf{q},\omega)$ in the momentum
range from $q=0$ to $q=2k_{F}$. Two types of contributions are clearly
visible: one is the collective Bogoliubov-Anderson phonon excitations
within the energy gap $\omega<E_{g}=2\Delta$ \cite{Combescot2006},
which exhibit themselves as a sharp $\delta$-peak in the structure
factor spectrum. Right above the energy gap,  a much broader distribution
emerges, which should be attributed to the fermionic single-particle
excitations by breaking Cooper pairs.

\begin{figure}
\begin{centering}
\includegraphics[width=0.45\textwidth]{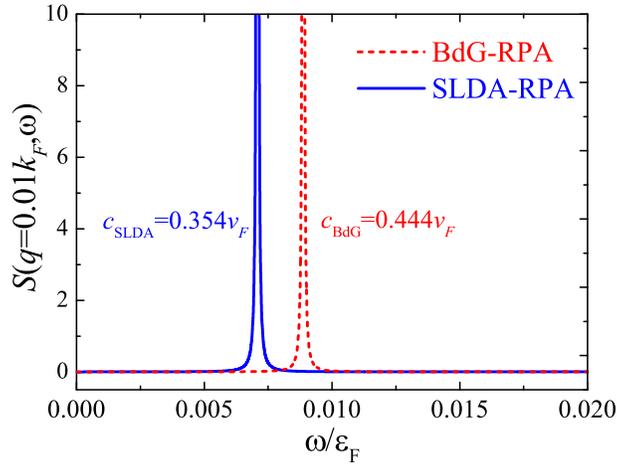} \caption{(color online) The phonon peak of the dynamic structure factor of
a unitary Fermi gas in the low-$q$ limit. The blue solid line is
our SLDA-RPA's prediction, while the red dashed line is the result
from the BdG-RPA theory. Here, to better represent the distribution
of a delta function at small $q$, a broadening width $\delta=10^{-4}\varepsilon_{F}$
has been used. The dynamic structure factor is measured in units of
$N/\varepsilon_{F}$.}
\par\end{centering}
\label{fig2} 
\end{figure}

\begin{figure}
\begin{centering}
\includegraphics[width=0.85\textwidth]{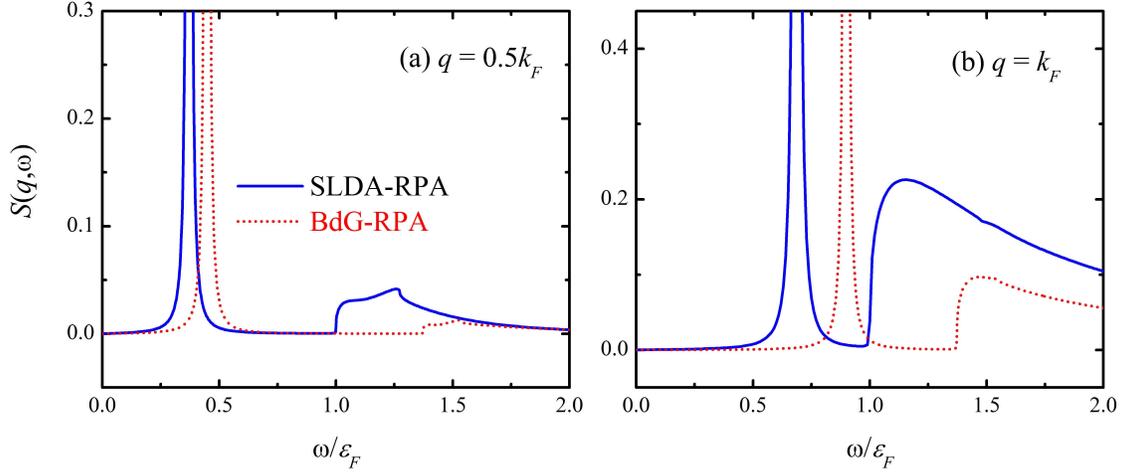} \caption{(color online) The dynamic structure factor of a unitary Fermi gas,
in units of $N/\varepsilon_{F}$, at $q=0.5k_{F}$ (a) and $q=k_{F}$
(b). The blue solid and red dashed lines show the results of the SLDA-RPA
and BdG-RPA theories, respectively. We note that, the scale for the
vertical axis in (a) and (b) is different.}
\par\end{centering}
\label{fig3} 
\end{figure}

A close examination of the phonon excitations is shown in Fig. 2 for
a very small transferred momentum $q=0.01k_{F}$. For comparison,
we also plot the result of the standard BdG-RPA prediction by a red
dashed line. It is anticipated that the dispersion of the phonon excitations
should follow $\omega=c_{s}q$, where $c_{s}$ is the sound velocity.
By fitting the position of the phonon peak as a function of $q$,
we numerically extract a value $c_{s}\simeq0.354v_{F}$, which coincides,
within the accuracy of our numerical calculations, with the value
obtained using the macroscopic definition of the sound speed, $c_{s}=\sqrt{(n/m)\partial\mu/\partial n}=\sqrt{\xi_{\mu}/3}v_{F}$.
This value is also consistent with the results determined from the
experiments and from the \emph{ab-initio} Monte Carlo calculations.
The agreement is not surprising, since the SLDA parameters have been
chosen to reproduce the known equation of state and hence the sound
speed. It is worth noting that a similar phonon peak is also predicted
by the BdG-RPA theory (i.e., using the BdG energy density functional).
However, the BdG-RPA theory predicts a sound speed $c_{s}\simeq0.444v_{F}$,
which is about $30\%$ larger than the above mentioned SLDA-RPA result.

At larger transferred momentum, i.e., $q\apprge0.5k_{F}$, the single-particle
excitations start to make a notable contribution to the dynamic structure
factor above the threshold $\omega=2\Delta=\varepsilon_{F}$, as shown
in Fig. 3. The sharp rise of the single-particle contribution at $\omega=2\Delta$
is unlikely to be destroyed by the possible residue interactions between
Cooper pairs and unpaired fermions, which is not accounted for in
our theory. Therefore, it could serve as a useful feature to experimentally
determine the pairing gap in the two-photon Bragg scattering experiments
\cite{Vale2016}. We also note that, compared with our SLDA-RPA results,
the BdG-RPA theory predicts a much weaker response of the single-particle
excitations at a larger threshold. This difference between the SLDA-
and BdG-RPA predictions could be easily resolved experimentally.

\begin{figure}
\begin{centering}
\includegraphics[width=0.5\textwidth]{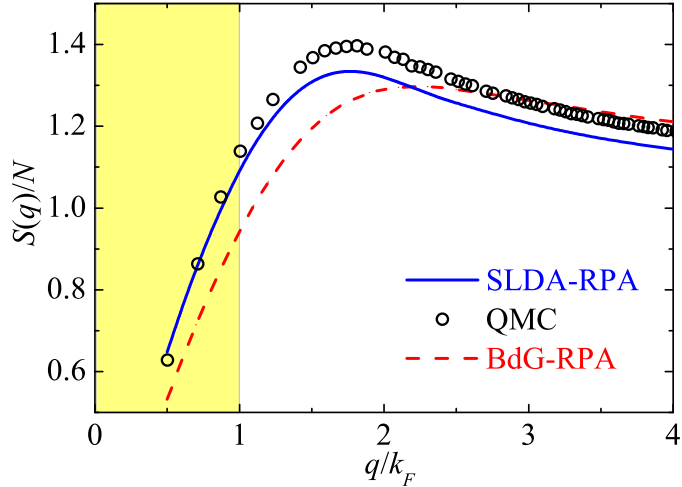} \caption{(color online) The static structure factor of a zero-temperature unitary
Fermi gas, calculated by the SLDA-RPA theory (blue solid line), in
comparison with the QMC result (black circles) \cite{Carlson2014}
and the BdG-RPA prediction (red dashed line). Our SLDA-RPA theory
is expected to be quantitatively reliable at $q\leq k_{F}$, as highlighted
by the yellow area.}
\par\end{centering}
\label{fig4} 
\end{figure}

A test of the accuracy of the theory can be obtained by looking at
the static structure factor

\begin{equation}
S(\mathbf{q})=\int d\omega S(\mathbf{q},\omega)\label{eq:ssf}
\end{equation}
for which for which QMC results are available \cite{Carlson2014,Combescot2006EPL}.
The comparison of our SLDA-RPA predictions with the latest diffusion
Monte Carlo data \cite{Carlson2014} is shown in Fig. 4, together
with the predictions of BdG-RPA. The excellent agreement between SLDA-RPA
and QMC at $q\le k_{F}$ is non-trivial and suggests that our theory
can be quantitatively reliable at small momentum transfer. Above the
Fermi momentum, instead, there are significant deviations. It is worth
noticing that the BdG-RPA theory gives results closer to QMC at large
momentum transfer, where the physics is dominated by single-particle
excitations and where BdG-RPA theory is known to work well \cite{Zou2010}.

\begin{figure}
\begin{centering}
\includegraphics[width=0.5\textwidth]{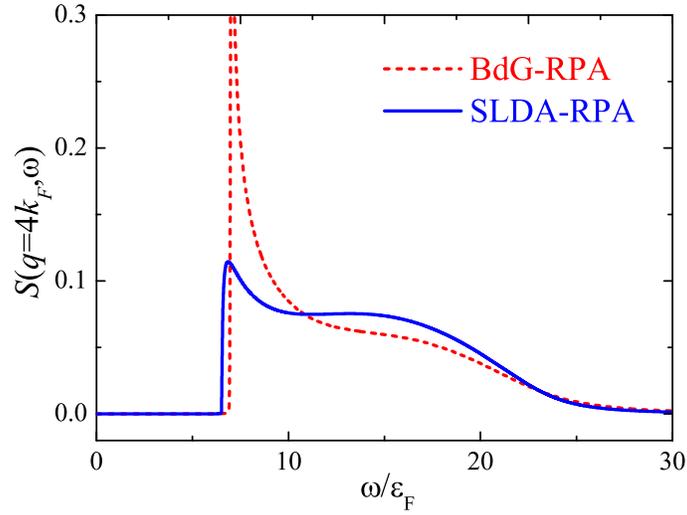} \caption{(color) The dynamic structure factor of a zero-temperature unitary
Fermi gas (in units of $N/\varepsilon_{F}$) at a large momentum transfer
$q=4k_{F}$, calculated by using the SLDA-RPA (blue line) and BdG-RPA
theories (red dashed line).}
\par\end{centering}
\label{fig5} 
\end{figure}

In Fig. 5, we show the dynamic structure factor at the momentum $q=4k_{F}$.
At such a large momentum, one can still separately resolve the bosonic
Cooper-pair excitations (i.e., a molecular peak structure at $\omega=q^{2}/4m=8\varepsilon_{F}$)
and fermionic single-particle excitations (i.e., the broader distribution
at $\omega=q^{2}/2m=16\varepsilon_{F}$). Compared with the BdG-RPA
result, our SLDA-RPA theory predicts a much smaller molecular peak.
This is understandable, since the SLDA theory is effectively a low-energy
theory and hence becomes less efficient at $\omega\gg\varepsilon_{F}$.
We note that, experimentally, there is a finite energy resolution
in the measurement of the dynamic structure factor \cite{Zou2010}.
The notable difference in the predictions for the molecular peak will
be easily smeared out by the finite energy resolution. As a result,
the SLDA-RPA approach may predict nearly the same line shape as the
BdG-RPA theory. The difference in the line shape is characterized
by the relative difference in the static structure factor, which is
about $5\%$. In the sense of predicting the experimental line shape
for the dynamic structure factor, we may argue that the SLDA-RPA is
\emph{semi-quantitatively} valid at large transferred momentum $q>k_{F}$. 

It should also be noted that an independent check of the SLDA-RPA
theory is provided by the $f$-sum rule \cite{Guo2010} 
\begin{equation}
\int d\omega\omega S(\mathbf{q},\omega)=\frac{N\mathbf{q}^{2}}{2m},\label{fsumslda}
\end{equation}
which should be satisfied. We have numerically checked that our SLDA-RPA
calculations obey this sum-rule within $1\%$ relative accuracy.

\section{Dynamic structure factor at the BCS-BEC crossover}

\label{cross}

In this section, we apply the SLDA-RPA theory to determine the dynamic
structure factor at the whole BCS-BEC crossover, by using the zero-temperature
chemical potential and pairing gap calculated from a Gaussian pair
fluctuation theory \cite{Hu2006} as the inputs. The energy density
functional Eq. (\ref{edslda}) - obtained under the scale invariance
assumption - is supposed to work well slightly away from the unitary
limit.

\begin{figure}[t]
\begin{centering}
\includegraphics[height=0.45\textwidth,angle=90]{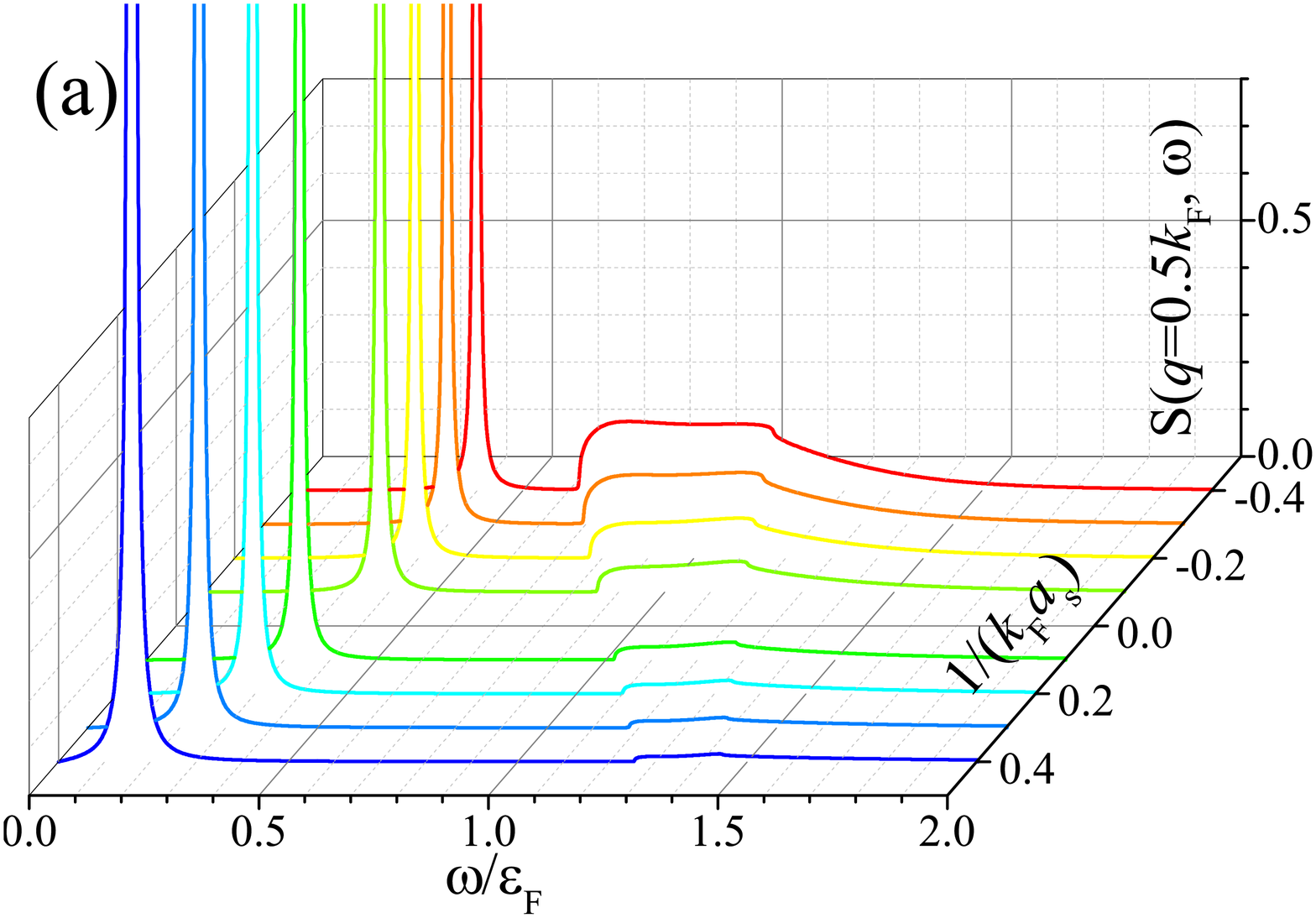}
\includegraphics[height=0.45\textwidth,angle=90]{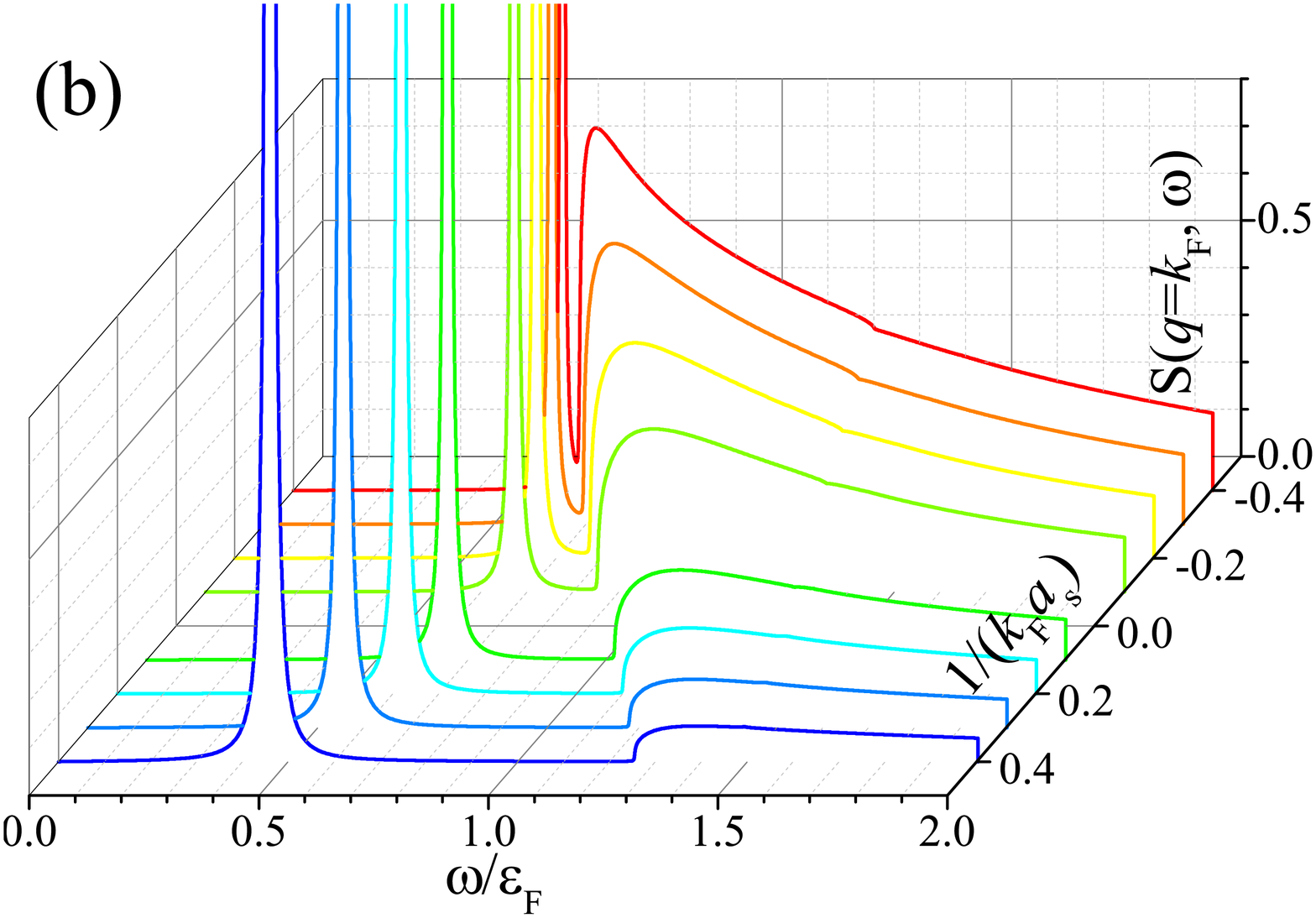}
\par\end{centering}
\begin{centering}
\caption{(color online) The dynamic structure factor (in units of $N/\varepsilon_{F}$)
at the BCS-BEC crossover and at the tansferred momentum $q=0.5k_{F}$
(a) and $q=k_{F}$ (b).}
\par\end{centering}
\label{fig6} 
\end{figure}

Fig. 6 reports the dynamic structure factor at the BCS-BEC crossover
at two different transferred momenta $q=0.5k_{F}$ (a) and
$q=k_{F}$ (b). On the BCS side, the single-particle contributions
become significant, as one may anticipate. Furthermore, at $q=k_{F}$
and $1/(k_{F}a)=-0.4$, where the bosonic peak position $\omega_{B}\sim c_{s}q$
is close to the two-particle scattering threshold $2\Delta$, there
is a strong overlap between the phonon and single-particle contributions,
leading to an interesting peak-dip-bump structure. When the system
crosses over to the BEC limit with increasing $1/(k_{F}a)$, the phonon
peak moves to the low energy, due to the decreasing sound velocity.
The single-particle contributions get suppressed very quickly. In
particular, at $q=0.5k_{F}$, the broader single-particle distribution
can be barely seen on the BEC side with $1/(k_{F}a)>0$.

Apparently, the experimental determination of the phonon peaks can
be ideally used to measure the sound velocity across the BCS-BEC crossover.
The measurement of the broader single-particle contributions may also
be useful to determine the pairing gap on the BCS side. 

\section{Conclusions}
\label{conclusions}

In summary, we have developed a random phase approximation theory
for calculating the dynamic structure factor of a strongly interacting
Fermi gas at unitarity and in the BCS-BEC crossover, within the framework
of a density functional theory approach \cite{Bulgac2007,Bulgac2013}.
The theory is expected to be quantitatively reliable at low transferred
momentum (i.e., $q<k_{F}$) and at low temperature (i.e., $T\ll T_{c}$),
where the predicted static structure factor agrees excellently well
with the result of the latest ab-initio diffusion quantum Monte Carlo
\cite{Carlson2014}. Therefore, our theory is useful to understand
the dynamic structure factor in the previously un-explored territory
of low transferred momentum, as schematically illustrated in Fig.
7 by a red rectangle. A stringent test of the applicability of our
theory could be obtained by comparing our predictions with the results
of on-going experiments \cite{Vale2016}.

\begin{figure}
\begin{centering}
\includegraphics[width=0.5\textwidth]{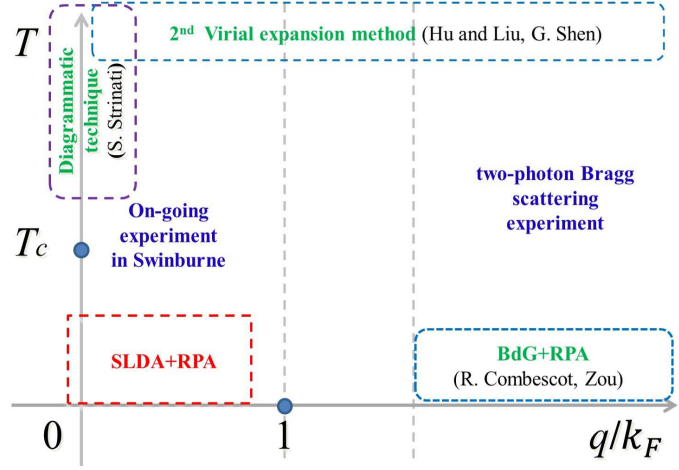} 
\caption{(color online) An illustration of the existing theories of the dynamic
structure factor of a strongly interacting Fermi gas, including the
virial expansion \cite{Hu2010PRA,Shen2013}, BdG-RPA theory \cite{Combescot2006,Combescot2006EPL,Zou2010}
and diagrammatic approach \cite{Palestini2012,He2016}. The applicable
parameter space of our SLDA-RPA theory is enclosed by the red dashed
line at small transferred momentum $q\leq k_{F}$ and at low temperature
$T\ll T_{c}$. The two-photon Bragg scattering experiment has so far
been carried out at $q\sim0.5k_{F}$ \cite{Vale2016} and $q\geq3k_{F}$
\cite{Veeravalli2008,Lingham2014}. The dashed borders of the domains
should not be considered as sharp boundaries, but just as an illustrative
guide.}
\par\end{centering}
\label{fig7} 
\end{figure}

\section{Acknowledgements}

We are grateful to Chris Vale, Sandro Stringari, Aurel Bulgac, Michael
McNeil Forbes and Lianyi He for fruitful discussions, and Stefano
Giorgini and Stefano Gandolfi for sharing their QMC data. PZ is indebted
to the BEC Center at Trento for hospitality when this work started.
This work was supported by the ARC Discovery Projects: FT130100815
and DP140103231 (HH), DP140100637 , and FT140100003 (XJL). RS acknowledges
support from DAE, Government of India. The work is also supported
by Provincia Autonoma di Trento (FD). Correspondence should be addressed
to PZ at phy.zoupeng@gmail.com.

\appendix
%dummy comment inserted by tex2lyx to ensure that this paragraph is not empty

\section{The response function $\chi^{0}$}

\label{app}

In this appendix, we discuss how to calculate the response function
$\chi^{0}$, by solving the stationary SLDA equation. The existence
of four different densities means that there will be 16 correlation
functions in $\chi^{0}$: 
\begin{equation}
\chi^{0}\equiv\left[\begin{array}{cccc}
\langle n_{1}n_{1}\rangle_{0} & \langle n_{1}n_{2}\rangle_{0} & \langle n_{1}n_{3}\rangle_{0} & \langle n_{1}n_{4}\rangle_{0}\\
\langle n_{2}n_{1}\rangle_{0} & \langle n_{2}n_{2}\rangle_{0} & \langle n_{2}n_{3}\rangle_{0} & \langle n_{2}n_{4}\rangle_{0}\\
\langle n_{3}n_{1}\rangle_{0} & \langle n_{3}n_{2}\rangle_{0} & \langle n_{3}n_{3}\rangle_{0} & \langle n_{3}n_{4}\rangle_{0}\\
\langle n_{4}n_{1}\rangle_{0} & \langle n_{4}n_{2}\rangle_{0} & \langle n_{4}n_{3}\rangle_{0} & \langle n_{4}n_{4}\rangle_{0}
\end{array}\right],\label{chi0def}
\end{equation}
where the abbreviation $\chi_{ij}^{0}=\langle n_{i}n_{j}\rangle_{0}$
is used. The derivation of these matrix elements is cumbersome. We
show here, as an example, the derivation of $\chi_{\uparrow\uparrow}^{0}\equiv\chi_{11}^{0}$.
According to the Wick theorem, and following the BCS theory, which
assume that only propagators - like $\langle\Psi_{\uparrow}^{\dag}\Psi_{\uparrow}\rangle$,
$\langle\Psi_{\downarrow}^{\dag}\Psi_{\downarrow}\rangle$, $\langle\Psi_{\downarrow}\Psi_{\uparrow}\rangle$
and $\langle\Psi_{\uparrow}^{\dag}\Psi_{\downarrow}^{\dag}\rangle$
- have a non-zero value, the imaginary-time Green's function $\chi_{11}^{0}(\textbf{r},\textbf{r}',\tau)\equiv-\left\langle T_{\tau}\left[\hat{n}_{1}(\textbf{r},\tau)\hat{n}_{1}(\textbf{r}',0)\right]\right\rangle $
can be written as 
\begin{equation}
\chi_{11}^{0}=-\left\langle \Psi_{\uparrow}^{\dagger}(\textbf{r},\tau)\Psi_{\uparrow}(\textbf{r}',0)\right\rangle \left\langle \Psi_{\uparrow}(\textbf{r},\tau)\Psi_{\uparrow}^{\dagger}(\textbf{r}',0)\right\rangle ,\label{chi11}
\end{equation}
where $\tau$ is the imaginary time and we assume $\tau>0$. By using
the Bogoliubov transformations 
\begin{equation}
\begin{split}\Psi_{\uparrow} & =\sum_{j}\left[u_{j\uparrow}(\textbf{r})c_{j\uparrow}e^{-iE_{j\uparrow}t}+v_{j\downarrow}^{*}(\textbf{r})c_{j\downarrow}^{\dag}e^{iE_{j\downarrow}t}\right],\\
\Psi_{\downarrow}^{\dag} & =\sum_{j}\left[u_{j\downarrow}^{*}(\textbf{r})c_{j\downarrow}^{\dag}e^{iE_{j\downarrow}t}-v_{j\uparrow}(\textbf{r})c_{j\uparrow}e^{-iE_{j\uparrow}t}\right],
\end{split}
\end{equation}
for the field operators $\Psi_{\sigma}$ and $\Psi_{\sigma}^{\dag}$,
one finds
\begin{equation}
\chi_{11}^{0}\left(\textbf{r},\textbf{r}',\tau\right)=-\sum_{i,j}u_{i}^{*}(\textbf{r})u_{i}(\textbf{r}')u_{j}(\textbf{r})u_{j}^{*}(\textbf{r}')f(E_{i})f(-E_{j})e^{(E_{i}-E_{j})\tau}.
\end{equation}
Here we use $\langle c_{i}^{\dagger}c_{j}\rangle=f(E_{i})\delta_{ij}$
and $\langle c_{i}c_{j}^{\dagger}\rangle=f(-E_{i})\delta_{ij}$, and
$f(x)=1/(e^{x/T}+1)$ is the Fermi distribution function of quasiparticles.
The spin index has been removed owing to the existence of a one-to-one
correspondence between the solutions of spin-up and spin-down energy
levels. By taking the Fourier transformation in the imaginary time,
$\chi_{11}^{0}(\textbf{r},\textbf{r}',i\nu_{n})=\int_{0}^{\beta}d\tau e^{i\nu_{n}\tau}\chi_{11}^{0}(\textbf{r},\textbf{r}',\tau)$,
where $\nu_{n}=2n\pi k_{B}T$ is the bosonic Matsubara frequency,
one obtains, 
\begin{equation}
\chi_{11}^{0}(\textbf{r},\textbf{r}',i\nu_{n})=\sum_{i,j}u_{i}^{*}(\textbf{r})u_{i}(\textbf{r}')u_{j}(\textbf{r})u_{j}^{*}(\textbf{r}')\frac{f(E_{i})-f(E_{j})}{i\nu_{n}+(E_{i}-E_{j})}.\label{resp_up_up_omega}
\end{equation}
For the homogeneous gas, a set of plane wave functions can be used
to expand the eigenfunctions $u_{i}$ in the form $u_{i}(\textbf{r})\rightarrow u_{k}e^{i\textbf{k}\textbf{r}}$.
By defining the transferring momentum $\textbf{p}=\textbf{k}'-\textbf{k}$
and the relative coordinate $\delta\textbf{r}=\textbf{r}-\textbf{r}'$,
then 
\begin{equation}
\chi_{11}^{0}(\delta\textbf{r},i\nu_{n})=\sum_{\mathbf{k},\mathbf{p}}\left|u_{\mathbf{k}}\right|^{2}\left|u_{\mathbf{k}+\mathbf{p}}\right|^{2}e^{i\textbf{p}\delta\textbf{r}}\frac{f(E_{\mathbf{k}})-f(E_{\mathbf{k}+\mathbf{p}})}{i\nu_{n}+(E_{\mathbf{k}}-E_{\mathbf{k}+\mathbf{p}})}.\label{resp_plane_back}
\end{equation}
By taking the Fourier transformation of the relative coordinate, $\chi_{11}^{0}(q,\omega_{n})=\int d\delta\textbf{r}\chi_{11}^{0}(\delta\textbf{r},i\omega_{n})e^{-i\textbf{q}\delta\textbf{r}}$,
we find that, 
\begin{equation}
\chi_{11}^{0}(\mathbf{q},i\nu_{n})=\sum_{\mathbf{k}}\left|u_{\mathbf{k}}\right|^{2}\left|u_{\mathbf{k}+\mathbf{p}}\right|^{2}\frac{f(E_{\mathbf{k}})-f(E_{\mathbf{k}+\mathbf{p}})}{i\nu_{n}+(E_{\mathbf{k}}-E_{\mathbf{k}+\mathbf{p}})}.\label{resp_q_omega}
\end{equation}
Using the expressions for $u_{\mathbf{k}}$ and $u_{\mathbf{k}+\mathbf{p}}$,
at zero temperature we obtain,

\begin{equation}
\chi_{11}^{0}(\mathbf{q},i\nu_{n})=\sum_{\mathbf{k}}\frac{1}{2}\left(1-\frac{\xi_{\mathbf{k}}\xi_{\mathbf{k}+\mathbf{q}}}{E_{\mathbf{k}}E_{\mathbf{k}+\mathbf{q}}}\right)\frac{E_{\mathbf{k}}+E_{\mathbf{k}+\mathbf{q}}}{\left(i\nu_{n}\right)^{2}-\left(E_{\mathbf{k}}+E_{\mathbf{k}+\mathbf{q}}\right)^{2}}.\label{resp_uu_inte-1}
\end{equation}
Through a similar process, we can derive the other 15 matrix elements
of $\chi^{0}$. In fact, after checking their expressions, only six
of them are independent. The remaining expressions are simply related
to each other by, for example, the replacement $\textbf{k}\rightarrow-\textbf{k}-\textbf{q}$.
In the following, we list the other five expressions for $\chi_{12}^{0}$,
$\chi_{13}^{0}$, $\chi_{14}^{0}$, $\chi_{34}^{0}$ and $\chi_{43}^{0}$
at zero temperature: 

\begin{eqnarray}
\chi_{12}^{0} & = & \sum_{\mathbf{k}}\frac{1}{2}\frac{\Delta^{2}}{E_{\mathbf{k}}E_{\mathbf{k}+\mathbf{q}}}\frac{E_{\mathbf{k}}+E_{\mathbf{k}+\mathbf{q}}}{(i\nu_{n})^{2}-(E_{\mathbf{k}}+E_{\mathbf{k}+\mathbf{q}})^{2}},\\
\chi_{13}^{0} & = & \sum_{\mathbf{k}}\frac{\Delta}{4}\left[\frac{\left(\xi_{\mathbf{k}}+\xi_{\mathbf{k}+\mathbf{q}}\right)}{E_{\mathbf{k}}E_{\mathbf{k}+\mathbf{q}}}\frac{E_{\mathbf{k}}+E_{\mathbf{k}+\mathbf{q}}}{(i\nu_{n})^{2}-(E_{\mathbf{k}}+E_{\mathbf{k}+\mathbf{q}})^{2}}-\left(\frac{1}{E_{\mathbf{k}}}+\frac{1}{E_{\mathbf{k}+\mathbf{q}}}\right)\frac{i\nu_{n}}{(i\nu_{n})^{2}-(E_{\mathbf{k}}+E_{\mathbf{k}+\mathbf{q}})^{2}}\right],\\
\chi_{14}^{0} & = & \sum_{\mathbf{k}}\frac{\Delta}{4}\left[\frac{\left(\xi_{\mathbf{k}}+\xi_{\mathbf{k}+\mathbf{q}}\right)}{E_{\mathbf{k}}E_{\mathbf{k}+\mathbf{q}}}\frac{E_{\mathbf{k}}+E_{\mathbf{k}+\mathbf{q}}}{(i\nu_{n})^{2}-(E_{\mathbf{k}}+E_{\mathbf{k}+\mathbf{q}})^{2}}+\left(\frac{1}{E_{\mathbf{k}}}+\frac{1}{E_{\mathbf{k}+\mathbf{q}}}\right)\frac{i\nu_{n}}{(i\nu_{n})^{2}-(E_{\mathbf{k}}+E_{\mathbf{k}+\mathbf{q}})^{2}}\right],\\
\chi_{34}^{0} & = & \sum_{\mathbf{k}}\frac{1}{2}\left[\left(1+\frac{\xi_{\mathbf{k}}}{E_{\mathbf{k}}}\frac{\xi_{\mathbf{k}+\mathbf{q}}}{E_{\mathbf{k}+\mathbf{q}}}\right)\frac{E_{\mathbf{k}}+E_{\mathbf{k}+\mathbf{q}}}{(i\nu_{n})^{2}-(E_{\mathbf{k}}+E_{\mathbf{k}+\mathbf{q}})^{2}}+\left(\frac{\xi_{\mathbf{k}}}{E_{\mathbf{k}}}+\frac{\xi_{\mathbf{k}+\mathbf{q}}}{E_{\mathbf{k}+\mathbf{q}}}\right)\frac{i\nu_{n}}{(i\nu_{n})^{2}-(E_{\mathbf{k}}+E_{\mathbf{k}+\mathbf{q}})^{2}}\right],\\
\chi_{43}^{0} & = & \sum_{\mathbf{k}}\frac{1}{2}\left[\left(1+\frac{\xi_{\mathbf{k}}}{E_{\mathbf{k}}}\frac{\xi_{\mathbf{k}+\mathbf{q}}}{E_{\mathbf{k}+\mathbf{q}}}\right)\frac{E_{\mathbf{k}}+E_{\mathbf{k}+\mathbf{q}}}{(i\nu_{n})^{2}-(E_{\mathbf{k}}+E_{\mathbf{k}+\mathbf{q}})^{2}}-\left(\frac{\xi_{\mathbf{k}}}{E_{\mathbf{k}}}+\frac{\xi_{\mathbf{k}+\mathbf{q}}}{E_{\mathbf{k}+\mathbf{q}}}\right)\frac{i\nu_{n}}{(i\nu_{n})^{2}-(E_{\mathbf{k}}+E_{\mathbf{k}+\mathbf{q}})^{2}}\right].
\end{eqnarray}
We note that, $\chi_{34}^{0}$ and $\chi_{43}^{0}$ should be regularized
in order to remove the ultraviolet divergence.


\begin{thebibliography}{10}
\bibitem{Bloch2008}I. Bloch, J. Dalibard, and W. Zwerger, Rev. Mod.
Phys. \textbf{80}, 885 (2008).

\bibitem{Leggett1980}A. J. Leggett, \textit{Diatomic molecules and
Cooper pairs}, in \textit{Modern Trends in the Theory of Condensed
Matter, Lecture Notes in Physics}, Vol. 115 (Springer-Verlag, Berlin,
1980).

\bibitem{Giorgini2008}S. Giorgini, L. P. Pitaevskii, and S. Stringari,
Rev. Mod. Phys. \textbf{80}, 1215 (2008).

\bibitem{Ho2004}T.-L. Ho, Phys. Rev. Lett. \textbf{92}, 090402 (2004).

\bibitem{Hu2007}H. Hu, P. D. Drummond, and X.-J. Liu, Nature Phys.
\textbf{3}, 469 (2007).

\bibitem{Hu2010}H. Hu, X.-J. Liu, and P. D. Drummond, New J. Phys.
\textbf{12}, 063038 (2010).

\bibitem{Astrakharchik2004}G. E. Astrakharchik, J. Boronat, J. Casulleras,
and S. Giorgini, Phys. Rev. Lett. \textbf{93}, 200404 (2004).

\bibitem{Bulgac2006}A. Bulgac, J. E. Drut, and P. Magierski, Phys.
Rev. Lett. \textbf{96}, 090404 (2006).

\bibitem{Burovski2008}E. Burovski, E. Kozik, N. Prokof'ev, B. Svistunov,
and M. Troyer, Phys. Rev. Lett. \textbf{101}, 090402 (2008).

\bibitem{Carlson2005}J. Carlson and S. Reddy, Phys. Rev. Lett. \textbf{95},
060401 (2005).

\bibitem{Carlson2008}J. Carlson and S. Reddy, Phys. Rev. Lett. \textbf{100},
150403 (2008).

\bibitem{Carlson2011}J. Carlson, S. Gandolfi, K. E. Schmidt, and
S. Zhang, Phys. Rev. A \textbf{84}, 061602 (2011).

\bibitem{Forbes2011}M. M. Forbes, S. Gandolfi, and A. Gezerlis, Phys.
Rev. Lett. \textbf{106}, 235303 (2011).

\bibitem{Gandolfi2014}S. Gandolfi, J. Phys.: Conf. Ser. \textbf{529},
012011 (2014).

\bibitem{Carlson2014}J. Carlson and S. Gandolfi, Phys. Rev. A \textbf{90},
011601(R) (2014).

\bibitem{Ohashi2003}Y. Ohashi and A. Griffin, Phys. Rev. Lett. \textbf{89},
130402 (2002); Phys. Rev. A \textbf{67}, 063612 (2003).

\bibitem{Liu2005}X.-J. Liu and H. Hu, Phys. Rev. A \textbf{72}, 063613
(2005).

\bibitem{Chen2005}Q. Chen, J. Stajic, S. Tan, and K. Levin, Phys.
Rep. \textbf{412}, 1 (2005).

\bibitem{Hu2006}H. Hu, X.-J. Liu, and P. D. Drummond, Europhys. Lett.
\textbf{74}, 574 (2006). 

\bibitem{Haussmann2007}R. Haussmann, W. Rantner, S. Cerrito, and
W. Zwerger, Phys. Rev. A \textbf{75}, 023610 (2007).

\bibitem{Diener2008}R. B. Diener, R. Sensarma, and M. Randeria, Phys.
Rev. A \textbf{77}, 023626 (2008).

\bibitem{Mulkerin2016}B. C. Mulkerin, X.-J. Liu, and H. Hu, Phys.
Rev. A \textbf{94}, 013610 (2016).

\bibitem{Luo2007}L. Luo, B. Clancy, J. Joseph, J. Kinast, and J.
E. Thomas, Phys. Rev. Lett. \textbf{98}, 080402 (2007).

\bibitem{Nascimbene2010} S. Nascimb\`ene, N. Navon, K. J. Jiang,
F. Chevy, and C. Salomon, Nature (London) \textbf{463}, 1057 (2010).

\bibitem{Horikoshi2010}M. Horikoshi, S. Nakajima, M. Ueda, and T.
Mukaiyama, Science \textbf{327}, 442 (2010).

\bibitem{Navon2010}N. Navon, S. Nascimb\`ene, F. Chevy, and C. Salomon,
Science \textbf{328}, 729 (2010).

\bibitem{Ku2012}M. J. H. Ku, A. T. Sommer, L.W. Cheuk, and M.W. Zwierlein,
Science \textbf{335}, 563 (2012).

\bibitem{Hu2004}H. Hu, A. Minguzzi, X.-J. Liu, and M. P. Tosi, Phys.
Rev. Lett. \textbf{93}, 190403 (2004). 

\bibitem{Altmeyer2007}A. Altmeyer, S. Riedl, C. Kohstall, M. J. Wright,
R. Geursen, M. Bartenstein, C. Chin, J. Hecker Denschlag, and R. Grimm,
Phys. Rev. Lett. \textbf{98}, 040401 (2007).

\bibitem{Schunck2007}C. H. Schunck, Y. Shin, A. Schirotzek, M. W.
Zwierlein, and W. Ketterle, Science \textbf{316}, 867 (2007).

\bibitem{Schirotzek2008}A. Schirotzek, Y. Shin, C.H. Schunck, and
W. Ketterle, Phys. Rev. Lett. \textbf{101}, 140403 (2008).

\bibitem{Massignan2008}P. Massignan, G. M. Bruun, and H. T. C. Stoof,
Phys. Rev. A \textbf{77}, 031601(R) (2008).

\bibitem{Chen2009}Q. J. Chen and K. Levin, Phys. Rev. Lett. \textbf{102},
190402 (2009).

\bibitem{Gaebler2010}J. P. Gaebler, J. T. Stewart, T. E. Drake, D.
S. Jin, A. Perali, P. Pieri, and G. C. Strinati, Nature Phys. \textbf{6},
569 (2010).

\bibitem{Hu2010PRL}H. Hu, X.-J. Liu, P. D. Drummond, and H. Dong,
Phys. Rev. Lett. \textbf{104}, 240407 (2010).

\bibitem{Combescot2006}R. Combescot, M. Y. Kagan, and S. Stringari,
Phys. Rev. A \textbf{74}, 042717 (2006).

\bibitem{Veeravalli2008}G. Veeravalli, E. Kuhnle, P. Dyke, and C.
J. Vale, Phys. Rev. Lett. \textbf{101}, 250403 (2008).

\bibitem{Hu2012}For a recent review, see, for example, H. Hu, Front.
Phys. \textbf{7}, 98 (2012).

\bibitem{Lingham2014}M. G. Lingham, K. Fenech, S. Hoinka, and C.
J. Vale, Phys. Rev. Lett. \textbf{112}, 100404 (2014).

\bibitem{Vale2016}C. J. Vale, \textit{Low-lying excitations in a
strongly interacting Fermi gas}, invited conference presentation at
ICAP 2016.

\bibitem{Pitaevskii2003}L. Pitaevskii and S. Stringari, \textit{Bose-Einstein
Condensation} (Oxford University Press, 2003). 

\bibitem{Liu2009PRL}X.-J. Liu, H. Hu, and P. D. Drummond, Phys. Rev.
Lett. \textbf{102}, 160401 (2009).

\bibitem{Liu2013}X.-J. Liu, Phys. Rep. \textbf{524}, 37 (2013).

\bibitem{Hu2010PRA}H. Hu, X.-J. Liu, and P. D. Drummond, Phys. Rev.
A \textbf{81}, 033630 (2010).

\bibitem{Shen2013}G. Shen, Phys. Rev. A \textbf{87}, 033612 (2013).

\bibitem{Son2010}D. T. Son and E. G. Thompson, Phys. Rev. A \textbf{81},
063634 (2010).

\bibitem{Hu2012PRA}H. Hu and X.-J. Liu, Phys. Rev. A \textbf{85},
023612 (2012).

\bibitem{Hu2010NJP}H. Hu, E. Taylor, X.-J. Liu, S. Stringari, and
A. Griffin, New J. Phys. \textbf{12}, 043040 (2010).

\bibitem{Guo2010}H. Guo, C.-C. Chien, and K. Levin, Phys. Rev. Lett.
\textbf{105}, 120401 (2010).

\bibitem{Palestini2012}F. Palestini, P. Pieri, and G. C. Strinati
Phys. Rev. Lett. \textbf{108}, 080401 (2012).

\bibitem{He2016}L. He, Ann. Phys. \textbf{373}, 470 (2016).

\bibitem{Combescot2006EPL}R. Combescot, S. Giorgini, and S. Stringari,
Europhys. Lett. \textbf{75}, 695 (2006).

\bibitem{Zou2010}P. Zou, E. D. Kuhnle, C. J. Vale, and H. Hu, Phys.
Rev. A \textbf{82}, 061605(R) (2010).

\bibitem{Guo2013}H. Guo, C.-C. Chien, and Y. He, J. Low Temp. Phys.
\textbf{172}, 5 (2013).

\bibitem{Bulgac2002}A. Bulgac, Phys. Rev. C \textbf{65}, 051305(R)
(2002).

\bibitem{Yu2003}Y. Yu and A. Bulgac, Phys. Rev. Lett. \textbf{90},
222501 (2003).

\bibitem{Bulgac2007}A. Bulgac, Phys. Rev. A \textbf{76}, 040502(R)
(2007).

\bibitem{Zwerger2012} W. Zwerger (ed.), \textit{The BCS-BEC Crossover
and the Unitary Fermi Gas,} Lecture Notes in Physics, Vol. 836 (Springer-Verlag,
Berlin, 2012).

\bibitem{Hohenberg1964}P. Hohenberg and W. Kohn, Phys. Rev. \textbf{136},
864 (1964).

\bibitem{Kohn1965}W. Kohn and L.J. Sham, Phys. Rev. \textbf{140},
1133 (1965).

\bibitem{Kohn1999}W. Kohn, Rev. Mod. Phys. \textbf{71}, 1253 (1999).

\bibitem{Bulgac2011}A. Bulgac, Y.-L. Luo, P. Magierski, K. J. Roche,
and Y. Yu, Science \textbf{332}, 1288 (2011).

\bibitem{Bulgac2013}A. Bulgac, Annu. Rev. of Nucl. Part. Sci. \textbf{63},
97 (2013).

\bibitem{Minguzzi2001}A. Minguzzi, G. Ferrari, and Y. Castin, Eur.
Phys. J. D \textbf{17}, 49 (2001).

\bibitem{Bruun2001}G. M. Bruun and B. R. Mottelson, Phys. Rev. Lett.
\textbf{87}, 270403 (2001).

\bibitem{Liu2004}X.-J. Liu, H. Hu, A. Minguzzi, and M. P. Tosi, Phys.
Rev. A \textbf{69}, 043605 (2004).

\bibitem{Stringari2009} S. Stringari, Phys. Rev. Lett. \textbf{102},
110406 (2009).

\bibitem{Forbes2014} M. M. Forbes and R. Sharma, Phys. Rev. A \textbf{90},
043638 (2014).
\end{thebibliography}
\end{document}